\documentclass[aps,prl,twocolumn,showpacs]{revtex4}
\usepackage{epsf}
 
\begin{document}

       \title{Quantum fluctuations of the ultracold atom-molecule mixtures}
      \author{T.\ Domanski}
\affiliation{Institute of Physics, 
             M.\ Curie Sk\l odowska University, 
	     20-031 Lublin, Poland}

\begin{abstract}
We investigate evolution of the quantum coherence in the ultracold
mixture of fermionic atoms and bosonic dimer molecules. Interactions 
are there experimentally controlled via tuning the external magnetic 
field. Consequently, the fermionic atoms and their bosonic counterparts 
can be driven to a behavior resembling the usual BCS to BEC crossover. 
We analyze in some detail how this quantum coherence evolves with 
respect to time upon a smooth and abrupt sweep across the Feshbach 
resonance inducing the atom-molecule quantum fluctuations.
\end{abstract}

\maketitle

\section{Atom superfluidity}

The recent experimental techniques for trapping and cooling 
of the atomic vapors enabled exploration of the extremely low 
temperature regions where quantum effects play a crucial role. 
An example can be the Bose Einstein condensation (BEC) produced out 
of the bosonic atoms in alkali metals, polarized hydrogen, etc. 
Phase transition to the BEC state is there triggered purely by 
the quantum statistical requirements which lead to macroscopic 
occupancy of the lowest energy level and can occur even in absence 
of any interactions. Recent activities in the field of ultracold 
atomic systems focus on application of similar techniques 
to fermionic atoms like $^{6}$Li or $^{40}$K (besides even 
number of nucleons they consist of odd number of electrons). 
At ultralow temperatures such quantum effects like the Pauli 
principle play considerable role, but eventual quantum phase 
transitions would be allowed only if fermionic atoms get 
correlated via interactions.

Interactions between trapped atoms are routinely induced 
by applying the magnetic field to fermion atoms prepared 
in several (two or more) hyperfine configurations. From 
elementary considerations \cite{Timmermans-99} it turns out 
that the involved hyperfine states experience the effective 
scattering described by a potential whose magnitude and sign 
depend on the applied field $B$. In particular, the various 
(so called) Feshbach resonances can take place. On this basis
there was proposed a mechanism of the resonance superfluidity 
\cite{Holland-01} with a transition occurring near the Fermi 
temperature $T_{c} \sim T_{F}$. Besides the isotropic phase 
there has already been observed also the exotic $p$-wave 
superfluidity \cite{Ticknor-04,p-wave}. 

A unique possibility of controlling the effective interactions
gives a chance for the experimental realization of the BCS 
to BEC crossover. The BCS limit corresponds to a case of weakly 
attracting fermion atoms which get coupled into the large Cooper 
pairs. In opposite limit, the tightly bound diatomic molecules 
are formed which ultimately can undergo transition to the BEC. 
Experimentalists are able to switch between these limits in 
a controllable manner and change of the interactions 
can be performed either adiabatically (by slowly varying the 
field) \cite{slow_sweep} or in a non-adiabatic way (via the 
sudden sweep) \cite{fast_sweep}.

In this short paper we investigate the quantum fluctuations 
induced by the time-dependent change of the interactions. We 
focus on a situation when the magnetic field is detuned from 
the resonant value $B_{0}$ towards the far BCS regime at 
higher field $B>B_{0}$. We consider two different processes: 
the smooth and the sudden switching. The fast sweep has been 
discussed in the literature but with some ambiguous conclusions
concernig evolution of the order parameters with respect 
to time \cite{Andreev-04,Barankov-04,Szymanska-05}. From our 
analysis we find that both parameters do oscillate 
in a damped way.

\section{Heisenberg equations}

In a close proximity to the Feshbach resonance (i.e.\ when 
$B \sim B_{0}$) the ultracold fermion atoms coexist and 
interact with the diatomic molecules. On a microscopic basis
this situation can be described in terms of the two component
boson fermion Hamiltonian \cite{Holland-01}
\begin{eqnarray}
H & = & \sum_{{\bf k},\sigma} (\varepsilon^{F}_{\bf k}-\mu) 
c_{{\bf k}\sigma}^{\dagger} c_{{\bf k}\sigma} 
+ \sum_{\bf q} \left( \varepsilon^{B}_{\bf q} \! + \! 
2\nu(B) \! - \! 2\mu \right) b_{\bf q}^{\dagger} b_{\bf q} 
\nonumber \\ & + &
\frac{g}{\sqrt{N}} \sum_{{\bf k},{\bf q}} \left(  
b_{\bf q}^{\dagger} c_{{\bf q}-{\bf k}\downarrow}
c_{{\bf k}\uparrow} + c_{{\bf k}\uparrow}^{\dagger}
c_{{\bf q}-{\bf k}\downarrow}^{\dagger} b_{\bf q}
\right) 
\label{BF}
\end{eqnarray}
which has been known and studied in the solid state 
physics by J.\ Ranninger and coworkers \cite{Micnas_etal} 
as a phenomenological model for the high temperature 
superconductivity. In the present context (\ref{BF}) 
describes the atoms in two hyperfine states denoted 
symbolically by $\sigma \! = \! \uparrow$ and $\downarrow$. 
The second quantization operators $c_{{\bf k}\sigma}^{(\dagger)}$, 
$b_{\bf q}^{(\dagger)}$ correspond to fermion atoms with energy 
$\varepsilon_{\bf k}^{F}=\hbar^{2}{\bf k}^{2}/2m$ and to diatomic
molecules with energy $\varepsilon_{\bf k}^{B}=\hbar^{2}
{\bf k}^{2}/2(2m)$. The effect of external magnetic field 
is included via the {\em detuning parameter} $\nu$ which 
shifts the boson energies and hence affects efficiency
of the boson-fermion coupling $g$ \cite{Domanski-pra03}. 
As usually $\mu$ is the common chemical potential and we 
use the grand canonical ensemble to ensure the conservation 
of the total particle number $\sum_{{\bf k},\sigma} c_{{\bf k}
\sigma}^{\dagger} c_{{\bf k}\sigma} + 2 \sum_{\bf q} 
b_{\bf q}^{\dagger} b_{\bf q}$. 

We are interested here in studying the time dependent 
evolution of fermion and boson occupancies together with
the corresponding order parameters. For this purpose
we derive the Heisenberg equations of motion which 
for the Hamiltonian (\ref{BF}) are given by
\begin{eqnarray}
i\frac{\partial c_{{\bf q}-{\bf k}\downarrow} c_{{\bf k}
\uparrow}}{\partial t} &  = & \left( \xi_{\bf k} + \xi_{
{\bf q}-{\bf k}} \right) c_{{\bf q}-{\bf k}\downarrow} 
c_{{\bf k}\uparrow} + g b_{\bf q} 
\label{H1} \\ & - &
g \sum_{\bf q'}
b_{\bf q'} \left( c_{{\bf q'}-{\bf k}\downarrow}^{\dagger} 
c_{{\bf q}-{\bf k}\downarrow} + c_{{\bf k}+{\bf q'}-{\bf q}
\uparrow}^{\dagger}c_{{\bf k}\uparrow} \right), \nonumber\\
i \sum_{\sigma} \frac{\partial  c_{{\bf k}\sigma}^{\dagger} 
c_{{\bf k}\sigma}}{\partial t} & = & 2 g \sum_{\bf q}
\left(b_{\bf q} c_{{\bf k}\uparrow}^{\dagger}
c_{{\bf q}-{\bf k}\downarrow}^{\dagger} - b_{\bf q}
^{\dagger} c_{{\bf q}-{\bf k}\downarrow}
c_{{\bf k}\uparrow} \right), \label{H2}\\
i\frac{\partial  b_{\bf q}}{\partial t}  & = & E_{\bf q} 
b_{\bf q} + g \sum_{\bf k} c_{{\bf q}-{\bf k}\downarrow} 
c_{{\bf k}\uparrow}, \label{H3}\\
i\frac{\partial b_{\bf q}^{\dagger} b_{\bf q}}{\partial t} 
& = & g \sum_{\bf k} \left( b_{\bf q}^{\dagger} c_{{\bf q}
-{\bf k}\downarrow} c_{{\bf k}\uparrow} - b_{\bf q} 
c_{{\bf k}\uparrow}^{\dagger} c_{{\bf q}-{\bf k}\downarrow}
^{\dagger} \right), \label{H4}
\end{eqnarray}
where $\xi_{\bf k}=\varepsilon^{F}_{\bf k}-\mu$, $E_{\bf q}
=\varepsilon^{B}+2\nu(B)-2\mu$ and we set $\hbar=1$. In
general equations (\ref{H1}-\ref{H4}) are not solvable exactly. 
In the next section we briefly discuss an approximate method 
which shall be valid for the ground state and for very low 
temperatures. 

\section{The single mode approach}

For temperatures close to the absolute zero we can neglect the 
excited (finite momentum) boson states. It is sufficient to 
restrict attention to the ${\bf q}\!=\!{\bf 0}$ boson level 
because it is macroscopically occupied. In such {\em single mode
approach} \cite{Andreev-04,Barankov-04} the initial Hamiltonian 
(\ref{BF}) reduces to
\begin{eqnarray}
H & = & \sum_{{\bf k},\sigma} \xi_{\bf k} 
c_{{\bf k}\sigma}^{\dagger} c_{{\bf k}\sigma} 
+ \sum_{\bf q} E_{\bf 0} b_{\bf 0}^{\dagger} b_{\bf 0} 
\nonumber \\
& + & \frac{g}{\sqrt{N}} \sum_{\bf k} \left(  
b_{\bf 0}^{\dagger} c_{-{\bf k}\downarrow}
c_{{\bf k}\uparrow} + c_{{\bf k}\uparrow}^{\dagger}
c_{-{\bf k}\downarrow}^{\dagger} b_{\bf 0}
\right) .
\label{single_mode}
\end{eqnarray}
Following Anderson \cite{Anderson-58} we introduce 
the pseudospin notation 
\begin{eqnarray}
\sigma_{\bf k}^{+} & \equiv  & 
c_{-{\bf k}\downarrow} c_{{\bf k}\uparrow},
\nonumber \\ 
\sigma_{\bf k}^{-} & \equiv & c_{{\bf k}\uparrow}^{\dagger}
c_{-{\bf k}\downarrow}^{\dagger}, 
\nonumber \\ 
\sigma_{\bf k}^{z} & \equiv & 1 - c_{{\bf k}\uparrow}^{\dagger}
c_{{\bf k}\uparrow} - c_{{\bf k}\downarrow}^{\dagger}
c_{{\bf k}\downarrow}
\nonumber
\end{eqnarray}
such that $\sigma_{\bf k}^{\pm}=\frac{1}{2}\left( \sigma_{\bf k}
^{x} \pm i \sigma_{\bf k}^{y} \right)$ and $\sigma_{\bf k}^{z}=
\left[\sigma_{\bf k}^{+},\sigma_{\bf k}^{-}\right]$ 
are the usual Pauli operators. In the single mode approach 
we rewrite the Heisenberg equations (\ref{H1}-\ref{H4}) using 
the pseudospin notation
\begin{eqnarray}
i \; \frac{\partial \sigma_{\bf k}^{+}}{\partial t} 
& = & 2 \xi_{\bf k}  \sigma_{\bf k}^{+} + g b_{\bf 0} 
\sigma_{\bf k}^{z}, \label{s1}\\
i \; \frac{\partial \sigma_{\bf k}^{z}}{\partial t} 
& = & 2 g \left(b_{\bf 0}^{\dagger} \sigma_{\bf k}^{+} 
- b_{\bf 0} \sigma_{\bf k}^{-} \right), 
\label{s2} \\
i \; \frac{\partial  b_{\bf 0}}{\partial t} & = & E_{\bf 0} 
b_{\bf 0} + g \sum_{\bf k} \sigma_{\bf k}^{+}, 
\label{s3} \\ 
i \; \frac{\partial b_{\bf 0}^{\dagger} b_{\bf 0}}
{\partial t} & = & g \sum_{\bf k} \left( b_{\bf 0}^{\dagger} 
\sigma_{\bf k}^{+} - b_{\bf 0} \sigma_{\bf k}^{-} \right)
\label{s4}
\end{eqnarray}
which are identical with expressions (5) and (6) in the Ref.\ 
\cite{Barankov-04}. One next replaces the boson operators by 
their time-dependent expectation values $\mbox{b}(t)=\langle 
b_{\bf 0}\rangle$ and $b^{*}(t)=\langle b_{\bf 0}^{\dagger}
\rangle$. 

In the stationary case when all parameters in (\ref{single_mode}) 
are time independent we can derive various expressions for the
static expectation values \cite{Micnas_etal}. Hamiltonian 
(\ref{single_mode}) has formally the following structure
\begin{eqnarray}
H = - \sum_{\bf k} \vec{h}_{\bf k} \cdot 
\vec{\sigma}_{\bf k} + \mbox{const}, \label{eq11}
\end{eqnarray} 
so the pseudospin $\vec{\sigma}_{\bf k}$ behaves as though 
affected by a fictitious magnetic field $\vec{h}_{\bf k}=
\left( -\Delta',\Delta'',\xi_{\bf k}\right)$ where
$\Delta' + i \Delta'' \equiv g \langle b_{\bf 0} \rangle$.
Following Anderson \cite{Anderson-58} we can solve this 
problem (\ref{eq11}) for arbitrary temperature. 
In analogy to the Weiss theory of ferromagnetism we obtain
that a magnitude of the pseudospin expectation value is 
$| \langle \vec{\sigma}_{\bf k} \rangle |= \mbox{tgh} \left\{ 
\sqrt{ \xi_{\bf k}^{2}+| \Delta |^{2} }/2k_{B}T \right\}$.
Determining the angle between the $z$ and $xy$ components of 
the vector $\vec{h}_{\bf k}$ we finally arrive at the stationary
equations \cite{Micnas_etal}
\begin{eqnarray}
\langle \sigma_{\bf k}^{+} \rangle & = & \langle c_{-{\bf k}
\downarrow} c_{{\bf k}\uparrow} \rangle
\nonumber \\
 &=&\frac{-\;g\Delta^{*}}
{ 2\sqrt{ \xi_{\bf k}^{2}+| \Delta |^{2} }} \; \mbox{tgh}\left\{ 
\frac{ \sqrt{ \xi_{\bf k}^{2}+| \Delta |^{2} }}{2k_{B}T}\right\}
\label{mean_field1}
\end{eqnarray}
and
\begin{eqnarray}
\langle \sigma_{\bf k}^{z} \rangle & = & 1-\sum_{\sigma} 
\langle c_{{\bf k}\sigma}^{\dagger} c_{{\bf k}\sigma} \rangle
\nonumber \\
 &=&  
 \frac{\xi_{\bf k}}{ \sqrt{ \xi_{\bf k}^{2}+| \Delta |^{2} }} \; 
 \mbox{tgh}\left\{ \frac{ \sqrt{ \xi_{\bf k}^{2}+| \Delta |^{2} }}
 {2k_{B}T}\right\}. \label{mean_field2}
\end{eqnarray}

\section{Fluctuations of the order parameters}

In the symmetry broken state (for $T\!<\!T_{c}$) the two component 
model (\ref{BF}) is characterized by two order parameters: $b(t)$ 
and another one of the fermion subsystem defined as $\chi(T)= 
\sum_{\bf k} \langle c_{-{\bf k}\downarrow} c_{{\bf k}\uparrow} 
\rangle$. These quantities are complex. In the stationary case 
they are proportional to each other as can be seen from the 
equation (\ref{mean_field1}). However, this relation is no 
longer valid when the Hamiltonian (\ref{single_mode}) depends 
on time. Evolution of the order parameters $b(t)$ and $\chi(t)$ 
with respect to time must be determined by solving the Heisenberg 
equations (\ref{s1}-\ref{s2}) subject to some boundary 
conditions. 

We analyze here such dynamics assuming that initially, for 
$t\!\leq\!0$, the mixed atom molecule system is exactly on 
the Fesbach resonance $\nu\!=\!\mu$. We also assume that 
the boson order parameter is real and $b(t\!\leq\!0)\!=1$ so
that the fermion order parameter is real too. Value of
$\chi(t\!\leq\!0)$ was determined from the equation 
(\ref{mean_field1}). For simplicity we focus on the ground 
state and set the boson fermion coupling $g$ as a unit or 
all the energies appearing in our study. 

\begin{figure}
\centerline{\epsfxsize=9cm \epsffile{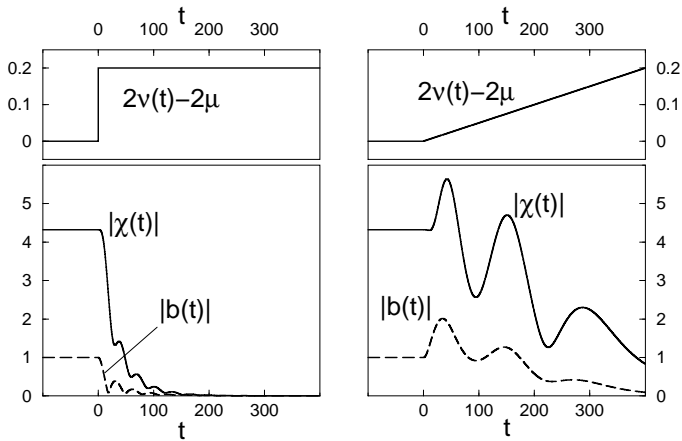}}
\caption{Variation of the fermion $|\chi(t)|$ and boson 
$|b(t)|$ order parameters caused by the detuning from 
the Feshbach resonance. Initially, for $t\!<\!0$, 
the system is in the stationary state with $\nu(t \! < \! 0)\! 
= \! \mu$ where the magnitudes of both order parameters are static. 
Upon lifting the molecule level there appear the damped
oscillations. In the left h.s.\ panel we illustrate 
the case of an abrupt detuning $\nu(t\!>\!0)\!=\!0.1g$ and in 
the right h.s.\  a smooth detuning $\nu(t) \propto t$.}
\label{Fig1}
\end{figure}

For time $t\!>\!0$ we change the detuning parameter $\nu$ in 
the following ways: a) via the sudden detuning as previously
discussed in the Refs  \cite{Andreev-04,Barankov-04,Szymanska-05} 
and b) through gradually increasing $\nu(t)-\mu \propto t$. 
Avoiding any constraint solutions we solved numerically 
the Heisenberg equations (\ref{s1},\ref{s2}) by means 
of the Runge-Kutta algorithm.

By increasing $\nu$ the boson fermion system is pushed to 
the far BCS regime. For $\nu\!=\!0.1 g$ at $t \rightarrow 
\infty$ the order parameters $b(t)$ and $\chi(t)$ should
reach very small values. Figure 1 shows that 
indeed temporal evolution occurs into such asymptotics and 
practically is achieved after several oscillations. In both 
cases the oscillations are clearly damped in agreement 
with the previous study by K.\ Burnet and coworkers 
\cite{Szymanska-05}. However, this process of damping 
is sensitive on a particular profile of the time dependent 
detuning. This can be seen from the figure 1 and also 
in the next figure 2, where we plot the phase $\theta$ 
of the boson order parameter $b(t)=|b(t)| e^{i\theta(t)}$. 
For a smooth switching the oscillations seem not to look 
regular at all.

\begin{figure}
\centerline{\epsfxsize=8cm \epsffile{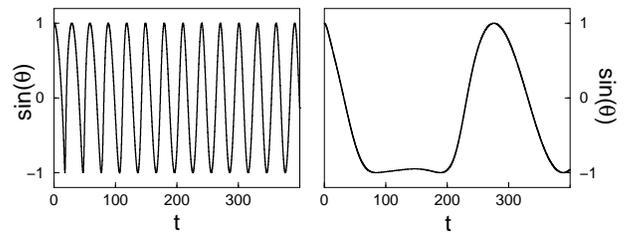}}
\caption{Evolution of the phase $\theta$ of the boson
order parameter $b(t)$ with respect to time $t$ for: 
an abrupt detuning (left), and for a smoothly increasing 
detuning $\nu(t)\propto t$ (right). Instead of the
bare angle $\theta$ we plot the function $\sin \theta 
= \mbox{Im} b(t)/|b(t)|$.}
\label{Fig2}
\end{figure}

\section{Summary}

We studied the dynamics of the ultracold fermion atoms 
upon the sudden and gradual detuning from the Feshbach 
resonance. Such situation can be experimentally realized 
by switching the external magnetic field from $B_{0}$ to 
the higher values. From the selfconsistent numerical solution 
of the equations of motion we find that the order parameters 
start oscillating with the amplitude decaying in time. 
Such damped oscillations depend on the specific form in 
which the detuning $\nu(t)$ is carried out. We also
remark that the quantum oscillations of the order parameters 
turn out to be damped even on the level of the single 
mode approach without taking into account scattering to 
the finite boson momenta.

\section{Acknowledgement}

This work is supported by the Polish Committee of Scientific 
Research under the grant No.\ 2P03B06225.

\end{document}